\documentstyle[12pt,epsfig]{article}
\begin{document}

\begin{center}
{\bf INFLUENCE OF BEAM DIVERGENCE AND CRYSTAL\\
MOSAIC STRUCTURE UPON PARAMETRIC X-RAY  \\
RADIATION CHARACTERISTICS \\
}
\end{center}
\vspace{5mm}
\begin{center}
{\large\sl  A.P. Potylitsin } \\
\end{center}
\begin{center}
{\small	       Nuclear Physics Institute,\\
               Tomsk Polytechnic University, \\
               pr. Lenina 2A, P.O.Box 25, \\
               634050 Tomsk, Russia \\
               e-mail: pap@phtd.tpu.edu.ru
}
\end{center}

\vspace{2mm}
\begin{abstract}  Based on kinematic model of parametric X-ray 
        radiation (PXR), an approach for calculation of PXR 
        characteristics (spectrum, intensity, polarization and yield)
        has been developed. The approach
	allows to take into account the	beam divergence, 
        mosaicity, aperture sizes, influence of $K$-edge, 
        etc. using a uniform technique.
\end{abstract}

\newpage

     1.	The existing theoretical models	[1-3] describe the process of

	parametric $X$-ray radiation (PXR) for the monodirected	charged
        particles beam interacting with sufficiently thin ideal crystals 
        whose multiple	scattering can be neglected. The real conditions
	of an experiment, however, are far from being ideal.
		
        To provide a quantitative comparison of the theoretical	and 
        experimental data we have to correctly take into account such 
        phenomena as beam divergence, mosaicity, detector's finite 
        aperture, effects of the $K$-edge absorption and some	other 
        factors.

	The authors of [4] developed a phenomenological approach 
        making it possible to approximately account  for the effects of 
       multiple	scattering of particles	in the target upon photon 
       angular distribution and PXR spectrum. They assumed that 
        multiple scattering leads to effective broadening of angular 
       distribution of virtual photons connected with incidence particle.
       This approach made it possible to describe the experimental 
       results obtained for thin perfect crystals. However, in the 
       experiments [5-7] there  was noted a significant discrepancy
	between	the data by the proposed model and those of the experiment.
	 The authors of [5] observed that angular distribution of PXR 
        became narrower with decreasing electron energy. 
        Experimental results in	[6,7] show that	the widths of orientational
	dependences (OD) of PXR yield are essentially less than the calculated
        values for the thick crystal target used. The authors of work [6] 
        proposed an approach where the processes of PXR emission and multiple
	scattering are independent (incoherent model). This approach gives 
        results close to the experiment and may be used for corrected
	account	of multiple scattering.

	In [8] it was shown theoretically that mosaicity of crystal
	target has no effect on  the total PXR intensity. Recently in [9]
	studied	experimentally were mosaic structure effects on the PXR
	yield and spectral characteristics. The	results  obtained for higher
        orders of PXR show disagreement with the model [4]. The	authors	
        of [10]	developed a method for the account of mosaicity based 
        on convolution of PXR angular distribution with effective 
        distribution including multiple	scattering and mosaicity 
        in the same manner. The effect of these factors must be different.
        The present work offers an approach providing a uniform 
        technique, within  the kinematic model, to calculate the PXR
        characteristics (spectrum, polarization, angular spread,
        photon yield into a finite aperture for any experimental 	
        environment, photon yield into an open cone, brightness,
        i.e. the PXR intensity per a solid angle unit) taking into
        account all the factors enumerated earlier.

     2.	As an assumed expression we chose the formula derived in 
       [11] as the most clear:
       $$
     {{dN}\over{dZ}} = {{{{\sum_\alpha\alpha\omega^3
     {\vert\chi_{\vec g}\vert}^2}
      d\Omega}}\over{{2\pi{\varepsilon}_0^{3/2}
      \beta(1-\sqrt{{\varepsilon}_0}
      \vec{\beta}\vec n)}}}\bigg\lbrack{(\sqrt{\varepsilon}_0 
      \omega\vec{\beta}-\vec g)\vec e_{\vec{k}\alpha}
      \over(\vec k_\bot+\vec g_\bot)^2+{\displaystyle\omega^2\over\beta^2}
      \{\gamma^{-2}+\beta^2(1-\epsilon_0)\}}\bigg\rbrack^2. \eqno(1)
        $$
	Here and later in the text use is made of the system of
	units $\hbar=m_e=c=1$.
        In Eq.(1) $\displaystyle{\varepsilon_{0}} = 1-{\omega^2_p/ \omega^2}$,
	$\omega_p$  is the plasma frequency,  
        $\displaystyle\vec\beta = {\beta\vec n_0}$
	is the initial particle	(electron) velocity,  $\vec n_0$,
	$\vec n$ are the unit vectors in the direction of the initial
	electron and the PXR photon (with the energy $\omega$
	and momentum $\vec k$), $\vec g$ is the reciprocal lattice vector,
	$\displaystyle\vec e_{\vec k\alpha}$	 are the  polarization 
        unit vectors, $\bot$ is the index denoting the projection of 
        vectors into 	the plane perpendicular	to  $ \vec n$. By  
        $\displaystyle \vert\chi_{\vec g}\vert$
	we denote here the following value:
        $$
       {\vert\chi_{\vec g}\vert}^2={\vert S(\vec g)\vert^2 
       \exp(-2W)\bigg\lbrack-{{\omega_p^2}\over{\omega^2}}
       {F(\vec g)\over z}\bigg\rbrack^2}. \eqno(2) 
       $$
        In Eq.(2), $\vert S(\vec g)\vert^2$ is the structure	factor,
	$\exp(-	2W)$ in	the Debye-Waller factor, $F(\vec g)$ is	the
	Fourier	component of the spatial distribution of electrons in
	the crystal atom, with	$F(0)=	z $, where  $z$ is the total
	number of electrons in the atom.

	   For the sake	of convenience let us introduce	the following
	coordinate systems :

	1) The	main system  $(x,y,z)$	where the z-axis is directed
	along the electron momentum, i.e. $\vec	n_0 ={\{0,0,1}\}$.
	The $y$  axis is normal to the diffraction plane, i.e. the vectors
        $\vec g, \vec n , \vec n_0$ are placed on the  $(xz)$  plane.

	2) The coordinate system indexed  $(g)$	 is related to the $\vec g$
	vector that is directed	along  $z_g$. The  $g$ system is rotated with
	respect	to the main system to an angle of 
        $\displaystyle -({\pi\over2}-\rm\theta_B)$
	around the y-axis. Here	$\rm\theta_B$ is used to denote the crystal
	alignment angle (Bragg	angle).

	3) The	$(d)$  index denotes the detector's system related to the
	emitted	photon,	which is rotated with respect to the main
	system to an angle of  $\theta_d = 2\rm\theta_B$ around the  $y$ -axis.
	The photon momentum in this system has the following components:
        $$
        {\vec k }={\omega{\vec	n}} = \omega\{n_{xd},n_{yd}, n_{zd}\}=
        \omega\{\sin\theta\cos\varphi, \sin\theta\sin \varphi, \cos\theta\}
        $$
	The polar angle	is measured from the Bragg direction coinciding
	with $z_d$, and the azimuthal	angle -	from the diffraction
	plane  $(x_d z_d)$. In the small angle	approximation
        $$
        \vec n =\{\theta\cos\varphi,   \theta\sin\varphi,
        1-{{\displaystyle\theta^2\over2}}\}=\{\theta_x,\theta_y,1-
        {\displaystyle\theta_x^2+\theta_y^2\over2}\}
        $$
       	Here and later $\theta_x,\theta_y$ point the component
	angles of the PXR photon emission in the $d$ system.

	For the	sake of	illustration we	write the transformation
	of the  $\vec g$  and  $\vec n$  vectors during transit into 
        the main system:
        $$
        \left \{
        \begin {array}{c}
        g_x=g_{xg}\sin\rm\theta_{B} - g_{zg}\cos\rm\theta_B \\
        g_y=g_{yg}\\
        g_z=g_{zg}\sin\rm\theta_B + g_{xg}\cos\rm\theta_B \\
        \end{array}
        \right.
        \hspace*{1cm}
        \left \{
       \begin{array}{c}
        n_x=n_{xd}\cos\theta_d+n_{zd}\sin\theta_d \\
          n_y=n_{yd}\\
          n_z=n_{zd}\cos\theta_d-n_{xd}\sin\theta_d \\
        \end{array}
        \right.\ 
         $$

         For an ideal crystal in the system selected we have
        $g_{xg}	= g_{yg}= 0$,\,\, \, $ g_{zg} = g $.
	Write down the expression (1) making evident the dependence
	of the photon emission angles (taking into account that
	$\omega_p^2 << \omega^2$ ):
        $$
        {{dN}\over{dZ}}={{\alpha\omega\vert\chi_{\vec g}
        \vert^2}\over{2\pi(1-{3\over 2}{{\displaystyle\omega_p^2}
        \over\displaystyle\omega^2})
        (1-{1\over 2\displaystyle\gamma^2})[1-(1-{\displaystyle\omega_p^2
        \over\displaystyle 2\omega^2})(1-{1\over2\displaystyle\gamma^2})
        \cos\theta_d]}}
        \Lambda d\Omega=
        $$
        $$ 
        = {{\alpha\omega\vert\chi_{\vec g}\vert^2}
        \over{2\pi(1-\cos\theta_d})}\Lambda d\Omega.\eqno(1a)
       $$	

	where the angular photon pattern is described by the following
	distribution:
       $$
       {\Lambda}={{\sum_\alpha\vert((1-{1\over2\displaystyle\gamma^2})\vec n_0
       -{\displaystyle\vec g\over\displaystyle\omega})\vec e_{\vec k\alpha}\vert^2}\over
       \bigg\lbrack{(\displaystyle{\vec k_\bot\over\displaystyle\omega}+ 
       {\displaystyle\vec g_\bot\over\displaystyle\omega})^2 +
       \gamma^{-2}+{\displaystyle\omega_p^2 \over{\displaystyle\omega^2}}}
       \bigg\rbrack^2}. \eqno(1b)
       $$

       The expressions (1a) and (1b) were derived using the 
      approximation $\gamma >> 1$. In Eqs. (1a), (1b) and later by
       	$ \omega$ we denote the PXR photon energy that is defined using
      the conservation laws and depends on the crystal alignment and 
      photon emission angles in the following manner [11]: 
      $$
      {\omega} ={\vec g\vec n_0\over{1\over \beta}-{\sqrt\epsilon_0}
      \vec n\vec n_0} \eqno(3) 
      $$

      Introduce the unit polarization vector: ${\vec e_{\vec k1}} = 
      {{\displaystyle\lbrack\vec n,\vec n_0\rbrack}
       \over{\displaystyle\vert\lbrack{\vec n,\vec n_0} \rbrack\vert}}$,
      $\displaystyle{\vec e_{\vec k2}} = \lbrack{\vec e_{\vec k1}, 
       \vec n}\rbrack.$	
      Now in the detector's system we obtain the following expressions
     for $\displaystyle \vec e_{\vec k\alpha}$:
     $$ 
     \vec e_{\vec k1} = \bigg\{{{n_{yd}}\over{\sin\theta_d\sqrt{1 + 
     2\cot\theta_dn_{xd}}}},  -{{{n_{xd}\cos\theta_d+\sin\theta_d}
     \over{\sin\theta_d\sqrt{1+2\cot\theta_{d}n_{xd}}}}}, 0\bigg\}
     $$
     $$
     \vec e_{\vec k2} = \bigg\{-{{{{n_{xd}\cos2\theta_d + 
     \sin\theta_d\cos\theta_d}} \over{\sin\theta_d\sqrt{1 + 
     2\cot\theta_dn_{xd}}}}}, - {{n_{yd}\cos\theta_d}
     \over{\sin\theta_d\sqrt{1+2\cot\theta_dn_{xd}}}},
     $$
     $$
     {{2n_{xd}\sin\theta_d\cos\theta_d+\sin^2\theta_d}\over{\sin\theta_d
     \sqrt{1+2\cot\theta_dn_{xd}}}}\bigg\}
     $$

     From the above follows:
      $\displaystyle{\vec n_0\vec e_{\vec k1}} = 0, \ \ 
      {\vec n_0\vec e_{\vec k2}} 
     = {\sin\theta_d\sqrt{1+2\cot\theta_dn_{xd}}}$,
    $$
    {\vec g\vec e_{\vec k1}}= - {{gn_{yd}\cos\rm\theta_B}
    \over{\sin\theta_d\sqrt{1 + 2\cot\theta_dn_{xd}}}}, \eqno(4)
    $$
    $$
    {\vec g\vec e_{\vec k2}} = {{g\lbrack\sin\theta_d
     \cos(\theta_d-\rm\theta_B)
     + n_{xd}\cos(2\theta_d-\rm\theta_B)\rbrack}
    \over{\sin\theta_d\sqrt{1+2\cot\theta_dn_{xd}}}}
    $$

    Substituting (4) into (1b) and summing with respect to polarization
   we can obtain the numerator in the following form 
   $$
   \sum_\alpha\bigg\vert\bigg((1-{{1\over{2\gamma^2}})\vec n_0 -{{\vec g}
   \over{\omega}})\vec e_{\vec k\alpha}}\bigg\vert^2
   = {{1\over{\sin^2\theta_d(1+2\cot\theta_dn_{xd})}}}\times
   $$
   $$
   \times\bigg\{{{{g^2}\over{\omega^2}}}n_{yd}^2\cos^2\rm\theta_B + 
   \bigg\lbrack(1-{{1} \over{2\gamma^2}})\sin^2\theta_d
   (1 + 2\cot\theta_dn_{xd}), 
   $$
   $$- {g\over\omega}(\sin^2\theta_d\cos(\theta_d - \rm\theta_B)
   + n_{xd}\cos(2\theta_D-\rm\theta_B)\bigg)\bigg\rbrack^2\bigg\}.\eqno(5)
   $$
   From Eq.(3) we can get the relation: 
   $$
   {g\over\omega} = {1\over\sin\theta _B}\bigg\lbrack1-
   \cos\theta_d + n_{xd}\sin\theta_d + 
   $$
   $$+ \cos\theta_d{n_{xd}^2 + n_{yd}^2\over2} + {1\over2\gamma^2} +
   {\omega_p^2\over2\omega^2}\cos\theta_d\bigg\rbrack. \eqno(3a)
   $$
   For Bragg direction ($n_{xd} = n_{yd} = 0$) from Eq.(3) we can,
   using the equity  $\displaystyle 1-\cos\theta_d =1 - \cos2\rm\theta_B = 
   2 \sin^2\rm\theta_B, $ 
   obtain:
   $$
    \rm\omega _B = {g\sin\rm\theta_B\over1-\cos\theta_d + 
    {\displaystyle1\over\displaystyle2\gamma^2} + {\displaystyle\omega_p^2
     \over\displaystyle\omega^2} \cos\theta_d} = {g\over2\sin\rm\theta_B}
    \bigg(1-{{\gamma^{-2} + {\displaystyle\omega_p^2
    \over\displaystyle\omega^2}
    \cos\theta_d}\over{\displaystyle 4\sin^2\rm\theta_B}}\bigg) \eqno(3b)
   $$
   For the geometry corresponding to large alignment angles \\
    $\displaystyle(\rm\theta_B\ >> \gamma^{-1}$,   
    $\displaystyle\omega_p / \omega )$,
   following  Eq.(3b) we have: $\rm\omega_B = {\displaystyle 
    g\over\displaystyle 2sin\rm\theta_B} $, which agrees with the Bragg 
    law for real  $X$-ray photon diffraction.
   For small deviation from the Bragg direction $(n_{xd},\ n_{yd} << 1)$
   we can obtain an expression simpler than a somewhat awkward Eq.5.
    Upon substituting Eq.(3a) into Eq.(5) leaving in the expansion
   the terms not higher that $\displaystyle n_{xd}^2 $, $ n_{yd}^2 $,
   we obtain:
   $$
   \sum_{\alpha}\bigg\vert\bigg((1-{1\over2\gamma^2})\vec n_0 -
   {\vec g\over\omega}\bigg)\vec e_{\vec k\alpha}\bigg\vert^2 =
   n_{xd} \gamma^{-2}\sin\theta_d\cos\theta_d+n_{xd}^2{\cos^2}2\rm\theta_B+
   $$ 
   $$
   + n_{yd}^2= \theta_x\gamma^{-2}\sin\theta_d\cos\theta_d + 
   \theta_{x}^2\cos^2 {2\rm\theta_B} + \theta_y^2 \eqno(6)
   $$
   The denomination in Eq.(1) is calculated in the main system where
   $$
   {\vec k_\bot\over\omega} =\bigg\{{n_{xd}\cos\theta_d-\sqrt{1-n_{xd}^2
   - n_{yd}^2}\sin\theta_d,\,\,\,  n_{yd}  ,\,\,\ 0 }\bigg\}
   $$
   $${\vec g_\bot\over\omega } = \bigg\{{ -{g\over\omega}\cos\rm\theta_B,\,\, 0,
   \,\, 0}\bigg\}
   $$
   Upon calculating in a similar approximation we get:
   $$
   \bigg\lbrack\bigg({\vec k_\bot\over\omega} + {\vec g_\bot\over\omega})^2
   +{1\over\beta^2\gamma^2} + 1- \epsilon_0\bigg\rbrack^2 =
   \bigg\lbrack{n^2_{xd}+ n^2_{yd} +\gamma^{-2}+
    {\displaystyle\omega^2_p\over\omega^2} \bigg\rbrack}^2 =
   $$
   $$
   = (\theta^2_x+ \theta^2_y + \theta^2_{ph})^2	\eqno(7)
   $$
   Here $ \theta_{ph} $ is used to indicate the angle 
   $\theta_{ph} = \sqrt{\gamma^{-2} + 
   {\omega^2_p / \omega^2}} $.
  Thus, the angular distribution of the PXR response with  respect 
  to the Bragg direction is described by the following expression:
  $$
  \Lambda(\theta_x,\theta_y) ={{\theta_x \gamma^{-2}
  \sin\theta_d\cos\theta_d + \theta^2_x \cos^2{2\rm\theta_B} + \theta^2_y }
  \over{\bigg\lbrack\theta^2_x + \theta^2_y + \gamma^{-2}+ 
   {\displaystyle\omega^2_p \over\displaystyle\omega^2}\bigg\rbrack^2}}
   \eqno(8)
  $$
  For an ultrarelativistic case, when  $\displaystyle\theta_x \sim 
  \gamma^{-2},$  the first summand in the numerator may be neglected.
  The expression thus obtained appears, as  to it is be expected,
  to agree with a well-known distribution [4]. In the expression (8),
  however, the summand linear with respect to $ \theta_x $ gives an
  asymmetric contribution into the PXR angular distribution in the 
  horizontal plane. This contribution increases with decreasing electron
  energy. It is this summand which determines the asymmetry of the PXR 
  orientation dependence measured in the experiment [12] for the 
  electron energy $ E_e = 25$ MeV.
  Using Eqs.(1a) and (8) we can  obtain the PXR intensities into 
  open cone around Bragg direction.
  Replacing  $n_{xd}$, $n_{yd} $ and $n_{zd}$ by their values in 
  the spherical coordinate system we get the following:
  $$
   \Lambda ={\gamma^{-2}\sin\theta\cos\varphi\sin\theta_d\cos\theta_d
  + \sin^2\theta(\cos^2{2\rm\theta_B\cos^2\varphi+\sin^2\varphi)}
  \over\bigg\lbrack\sin^2\theta +\gamma^{-2}+{\displaystyle\omega^2_p\over
  \displaystyle\omega^2}\bigg\rbrack^2}
  $$
  which  is readily integrated, 
  $$
  \Upsilon =\int\limits^\pi_0\sin\theta{d\theta}\int\limits^{2\pi}_0{d\varphi}
  \Lambda(\theta,\varphi)=\pi(1+\cos^2{2\rm\theta_B})
  \bigg\lbrack\ln{2\over\displaystyle\theta^2_{ph}}
   -1\bigg\rbrack
   $$
   Upon integrating Eq.(1a) with respect to the crystal thickness
   $L$ (taking into  account the absorption) we obtain the
   well-known expression [3]:
   $$ 
   N_0\approx{{\alpha\rm\omega_B(1+\cos^2{2\rm\theta_B})\vert\chi_{\vec g}\vert^2}
   \over{2(1-\cos 2\rm\theta_B)}}\bigg(\ln{2\over\displaystyle\theta^2_{ph}}
   -1\bigg) L_a\bigg(1- \exp\big({- \frac{L}{L_a}}\big)\bigg) \eqno(9)
   $$
   Where $L_a$ is the absorption length of photon with the energy
   $\rm\omega _B$.

  3. In order to make account of real experimental conditions
  (beam divergence, mosaicity, finite aperture of the detector,
   etc.) we propose a simple algorithm. Let the beam divergence be 
  described by the distribution $\displaystyle F_e(\Delta_x,\Delta_y)$ 
  and the mosaic structure by  $\displaystyle F_m(\alpha_x,\alpha_y )$.
   Using the approximations
  $\Delta_{x,y} << 1$,  $\alpha_{x,y} << 1$, which are almost always 
  true, we may assume the PXR angular distribution to be invariable 
  (i.e. $\displaystyle\rm\omega_B, \  \displaystyle\theta_{ph}=\rm const)$.
  Changes occur only in the Bragg direction which is used to 
  determine the angles $\theta_x$,  $\theta_y$ in Eq.(8). It can be 
  demonstrated	that for the  electrons with the incidence angles 
  $\Delta_x$,  $\Delta_y$ (determined with respect to the mean 
  direction $\displaystyle < \vec n_0  >$) shift of the Bragg direction in the $d$ 
  - system is found using the following:
  $$
   n^{\rm B}_{xd} = -\cos\Delta_y\sin\Delta_x\approx-\Delta_x
  $$
  $$ 
  n^{\rm B}_{yd} = \sin\Delta_y\approx\Delta_y \eqno(10a)
  $$
  $$ 
  \Delta^{\rm B}_{zd}=\cos\Delta_y\cos\Delta_x\approx1-{\Delta^2_x +\Delta^2_y
  \over2}
  $$
  The mosaic distribution function  $\displaystyle F_m(\alpha_{x}, 
  \alpha_y)$ is defined with respect to the mean direction of 
  $< {\vec g>}$, i.e. 
  in the $g$-system. The Bragg direction for an element of mosaic 
  structure corresponding to the reciprocal lattice vector 
  $\displaystyle \vec g_g = g\{\alpha_x, \alpha_y, 
   1-{\alpha^2_x + \alpha^2_y\over2}\}$   will be determined as 
  $$
  n^{\rm B}_{xd}=-\sin2\rm\theta_B\sin^2\alpha_y + \sin2\alpha_x\cos^2\alpha_y
  \approx2\alpha_x
  $$
  $$
  n^{\rm B}_{yd} = -\sin2\alpha_y\sin(\rm\theta_B+\alpha_x)\approx-
  2\alpha_y\sin\rm\theta_B \eqno(10b)
  $$
  $$
  n^{\rm B}_{zd} = \cos2\rm\theta_B\sin^2\alpha_y + \cos^2\alpha_y\cos2\alpha_x
  \approx 1-2\alpha^2_x-2\alpha^2_y\sin^2\rm\theta_B
  $$

 Thus, for PXR generation by a diverging electron beam, the photon 
 angular spread with respect to the mean Bragg direction can be
 written using the convolution:
  $$
  \Lambda_e(\theta_x,\theta_y)=\int{d}\Delta_{x}d\Delta_yF_e(\Delta_x,
  \Delta_y) \Lambda(\theta_x+\Delta_x,\theta_y-\Delta_y) \eqno(11)
  $$
  If, alongside with the diverging beam, we have a mosaic crystal then we
  get an angular spread of the form: 
  $$
  \Lambda_{e,m}(\theta_x,\theta_y)=\int{d}\alpha_xd\alpha_yF_m(\alpha_x,
  \alpha_y)\Lambda_e(\theta_x-2\alpha_x,\theta_y + 2\alpha_y\sin\rm\theta_B)=
  $$
  $$ 
  =\int{d}\alpha_xd\alpha_yF_m(\alpha_x,\alpha_y)\int{d} \Delta_{x}d
  \Delta_yF_e(\Delta_x,\Delta_y)\times
  $$
  $$
  \times\Lambda(\theta_x+\Delta_x-2\alpha_x,\theta_y-\Delta_y+
  2\alpha_y\sin\rm\theta_B) \eqno(12)
   $$
 In order to provide the PXR yield into a finite detector's aperture
 Eq.(12) should	be integrated with respect to the aperture 
  $\Delta\Omega$:
 $$
  N_{\rm PXR}=\rm const\int\limits_{\Delta\Omega}d\theta_x 
   d\theta_y\Lambda_{e,m}(\theta_x,\theta_y)
  $$
 This expression could be simplified by introducing the variables \\
 $\displaystyle \xi_x = \Delta_x{-}2\alpha_x,$   
 $\displaystyle \xi_y=-\Delta_y+2\alpha_y\sin\theta_B$.
 Then the internal integral in Eq.(12) will have the form:
 $$
  -\int{d}\xi_{x}d\xi_yF_e(\xi_x+2\alpha_x,-\xi_y+2\alpha_y\sin\rm\theta_B)
  \Lambda(\theta_x+\xi_x,\theta_y+\xi_y)
  $$
  and the integral is transformed into:
  $$
  \Lambda_{e,m}(\theta_x,\theta_y)= -\int\int{d}\alpha_x{d}\alpha_y{F_m}
  (\alpha_x, \alpha_y)d\xi_{x}d\xi_{y}\times
  $$
  $$
  \times F_e(\xi_x+2\alpha_x, -\xi_y+2\alpha_y\sin\rm\theta_B)
   \Lambda(\theta_x+\xi_x,  \theta_y+\xi_y)
  $$
  If the effective angular distribution function is introduced, then
  $$
  F_{eff}(\xi_x,\xi_y)=	-\int{d}\alpha_x{d}\alpha_y{F_m}(\alpha_x,\alpha_y)
   F_e(\xi_x+2\alpha_x,	-\xi_y+2\alpha_y\sin\rm\theta_B), \eqno(13)
  $$
  which in a number of cases can be analytically calculated (e.g. 
  when  $F_m$ and  $F_e$ are Gaussian distributions), then instead
  of Eq.(12) we get:
  $$ 
  \Lambda_{e,m}(\theta_x, \ \theta_y)=\int{d}\xi_x{d}\xi_yF_{eff}(\xi_x,\xi_y)
  \Lambda(\theta_x+\xi_x,\theta_y+\xi_y) \eqno(14)
  $$
  For illustrative purposes let us calculate the effective angular 
 distribution  $\displaystyle F_{eff}(\xi_x,$   
  $\displaystyle \xi_y)$ when $\displaystyle F_m(\alpha_x, \ \alpha_y)$
  and $\displaystyle F_e(\Delta_x,  \Delta_y)$ are approximated by the 
  Gaussians :
  $$
  F_m(\alpha_x,\alpha_y)=\rm C_1 \rm exp\{-{\alpha^2_x\over
  2\displaystyle\sigma^2_m}\}
  \rm exp\{-{\alpha^2_y\over2\displaystyle \sigma^2_m}\},
  $$
  $$
  F_e(\Delta_x,\Delta_y)= \rm C_2 \rm exp\{-{\Delta^2_x
   \over2\displaystyle \sigma^2_x}\}
   \rm exp\{-{\Delta^2_y\over2\displaystyle\sigma^2_y}\}
  $$
 The latter approximation can describe the electron beam angular 
 divergence with different dispersion along $x$	and $y$.In this	
 case the effective angular distribution, Eq.(13), is readily 
 calculated:
 $$
  F_{eff}(\xi_x,\xi_y)= \rm C_3 \exp\big\{-{\xi^2_x\over2\sigma^2_x
  (1+4{\sigma^2_m/\sigma^2_x})}\big\} \exp
  \{-{\xi^2_y\over2\sigma^2_y (1+4{\sin^2\rm\theta_B\sigma^2_m/\sigma^2_y})}\}
  $$
 This results in broadened Gaussians with the dispersions
 $\displaystyle \sqrt{\sigma^2_x+ 4\sigma^2_m}$ and  \\
  $ \sqrt{\sigma^2_y+ 4\sigma^2_m \sin^2\rm\theta_B}$.

   The effect of the mosaic structure along the $x$-axis is four
  times that of the divergence.

 4. Let us consider in a greater detail the effects of divergence 
  and mosaic structure on PXR angular distribution. For the sake of
 convenience we consider one - dimensional distribution along 
 the $y$ axis.
	
   a) Let the beam divergence be described by the following
  distribution: 
  $$
  F_e(\Delta_x,\Delta_y)={\displaystyle 1 \over 2\pi\sigma^2_e}
  \exp(-{\Delta^2_x +\Delta^2_y
  \over2\sigma^2_e})= F_e(\Delta_x)F_e(\Delta_y), 
  $$
   with $\int{d}\Delta_xF_{e}(\Delta_x)=\int{d}
  \Delta_yF_{e}(\Delta_y)=1$

   Then the resulting angular distribution of PXR in vertical
  direction has the form: 
  $$
  \widetilde {\Lambda}(\theta_y)=\int{d}\theta_x\int{d}\Delta_x{d}
  \Delta_yF_e(\Delta_x)F_e(\Delta_y)\Lambda(\theta_x+\Delta_x,
  \theta_y-\Delta_y)
  $$
  If we change the sequence of integration in this expression
  we may analy \\tically calculate the integral
  $$
  \int{d}\theta_x\Lambda(\theta_x+\Delta_x, \theta_y-\Delta_y)
  \simeq\int{d\theta_x}\Lambda(\theta_x, \theta_y-\Delta_y), 
  $$
  since the integration limits can approach $\pm\infty$.
  As a result we obtain
  $$
  \Lambda(\theta_y-\Delta_y)= \int{d}\theta_x\Lambda(\theta_x,
   \theta_y-\Delta_y)=
  $$
  $$
  ={\pi\over2} {{\theta^2_{ph}\cos^2\rm\theta_B+(\theta_y
  -\Delta_y)^2 (1+\cos^{2}2\rm\theta_B)}\over\displaystyle[\theta^2_{ph}+
  (\theta_y-\Delta_y)^2]^{3\over2}} \eqno(15)
  $$
  Now integration with respect to $\displaystyle d\Delta_x$ becomes trivial.
  Thus 
  $$
  \widetilde \Lambda(\theta_y)= \int{d}\Delta_{y}F_e(\Delta_y)
  \Lambda(\theta_y-\Delta_y) \eqno(16)
  $$
  Consider the case with  $\displaystyle\rm\theta_B =\pi/4$. Shown in 
  Fig.1 curve 1 is the distribution of Eq.(8) for ideal case 
  ($\Delta_y =0$ )  versus a dimensionless variable 
  $y = \displaystyle{\theta_y\over\theta_{ph}}$. One may see a double-lobe 
  distribution	with maxima at \\ $\displaystyle y_0 =\pm\sqrt2$.	
  As follows from the  expressions obtained, the effects of beam 
  divergence (and, therefore those, of multiple scattering) are reduced
  to the  broadening maxima, the  decrease of the dip of minimum, however, 
  producing small  effect on the position of the maxima.  

  In Fig.1 curve 2 shows the convolution of the distribution (16)
  with the Gaussian $\displaystyle F_e(\Delta_y)$ for the dispersion 
  $\displaystyle \sigma_e=\theta_{ph}$.One may notice a slight 
  shift of the maxima  towards the region of large values. 
  Nevertheless, the derived  value of  $y_0$ is by far lower than 
 it follows from a well known   model [4]:
  $$
  y_0< \sqrt{\theta^2_{ph} + \sigma^2_e} /\theta_{ph}.
  $$
  It should be noted that the value of distribution contrast
  $\displaystyle \widetilde{\Lambda}(\theta_y)$ (the ratio of the 
  intensity at its   maximum and minimum ) can be used to define
 the angular distribution   of the initial beam.

  b) Mosaic structure effects could be considered in a similar manner.	
  Convolution of the Gaussian with the dispersion 
  $\displaystyle\sigma^2_m$ with the expression (15), was shown 
  in Fig.2, where
  $$
  \theta_y^\prime=\theta_y+2\alpha_y\sin\rm\theta_B
  $$
  Very roughly one can estimate the value of dispersion
   $\displaystyle\sigma^2_m$  where in the distribution the two maxima 
   are smoothed and a   single peak appears at $\displaystyle \theta_y= 0$; \
    $\displaystyle 1.18\sigma_m > \displaystyle\sqrt{2}\theta_{ph}
  \sin\rm\theta_B$,  or
   $\displaystyle\sigma_m > 1.25\theta_{ph}\sin\rm\theta_B$.

  Drawn in Fig.2 are the results of convolution of the exact
  distribution for different dispersions 
  $\displaystyle\sigma_m=\theta_{ph}$ and  $2\theta_{ph}$. As follows 
  from the figure the double - lobe distribution of PXR virtually disappears.

  5. It follows from Eq.(3) that the shape and width of the PXR	 
  spectral line are defined by the electron beam divergence, crystal 
  mosaic structure and the collimator aperture. In order to obtain
  the shape of spectral line we have to introduce the variable 
  $\displaystyle \theta_x=\theta_x(\omega)$ and then integrate 
  the expression with   respect to the remaining variable 
   $\displaystyle \theta_y$. To make a   simultaneous account of the 
   effects of divergence and mosaic 
  structure let us obtain the relation between the PXR photon energy 
  and the angles used in the problem. Substituting values of the 
  vectors  $\displaystyle \vec g, \vec n$, and $\displaystyle \vec n_0$ 
  in the main system into Eq.(3) we get the sought - for dependence 
  for the energy of PXR  photons emitting in the direction given
  $\displaystyle n_{xd}, n_{yd}$ in a 
  mosaic crystal for the electrons of a diverging beam:
  $$
  \omega={g\{\alpha_x\Delta_{x}\sin\rm\theta_B - \Delta_{x}\cos\rm\theta_B+
  \alpha_y\Delta_y + \alpha_{x}\cos\rm\theta_B + \sin\rm\theta_B}\times
  $$
  $$
  \times(1-{\alpha^2_x+\alpha^2_y+\Delta^2_x+\Delta^2_y\over2})\}/
   \bigg\lbrack1-\Delta_x(n_{xd}\cos\theta_d+\sin\theta_d)-
  $$
  $$-\Delta_{y}n_{yd}+n_{xd}\sin\theta_d-\cos\theta_d+\cos\theta_{d}\times
   {1\over2}(\Delta^2_x+\Delta^2_y+n^2_{xd}+n^2_{yd})\bigg\rbrack
  $$
  In the above equation we left the second-order terms. In a more 
   rough approximation, leaving the terms up to the first order, we 
   have:
   $$
   \omega={{g\sin\rm\theta_B\{1+(\alpha_x-\Delta_x)\cot\rm\theta_B\}}
   \over{(1-\cos\theta_d)\{1+{\displaystyle\sin\theta_d\over
   {\displaystyle 1-\cos\theta_d}}
   (n_{xd}-\Delta_x)\}}}
   $$

    For the geometry chosen $\displaystyle \theta_d=2\rm\theta_B$,  therefore
   $$
   \omega=\rm\omega_B\{1+(\alpha_x-\Delta_x)\cot\rm\theta_B-(n_{xd}-\Delta_x)
   \cot\rm\theta_B\}=
   $$
   $$=\rm\omega_B\{1+(\alpha_x-n_{xd})\cot\rm\theta_B\}=
   \rm\omega_B\{1+(\alpha_x-\theta_x)\cot\rm\theta_B\} \eqno(17)
   $$
   Here one may use the new spectral variable $\displaystyle u =
   \frac{\omega - \rm\omega_B}{\displaystyle\rm\omega_B}\tan\rm\theta_B =
   \alpha_x-\theta_x$.

   As is clear from the above the photon energy depends on the 
   mosaic structure in the diffraction plane and the photon exit
   angle only, and is irrespective of the electron beam divergence.
   From Eq.(17), for $\displaystyle \alpha_x=0$, we may obtain
   $$
    \displaystyle\omega =\rm\omega_B(1-\theta_{x}\cot\rm\theta_B); \ \
   \displaystyle\theta_x = - \frac{\displaystyle\omega -\rm\omega_B}
   {\displaystyle\rm\omega_B} \tan\rm\theta_B; \ \
   \displaystyle d\theta_x = - \frac{\displaystyle d\omega}
   {\displaystyle\rm\omega_B} \tan\rm\theta_B.
   $$

    Let us find the shape of a PXR spectral line for round
   aperture $\displaystyle \theta^2_{x}+\theta^2_{y} \le\theta^2_c$
   aligned along the Bragg direction. For this purpose we have
   to integrate the distribution (8) with respect to the angle
   $\theta_y$ within the aperture:
   $$
   \Lambda_c(\theta_x) =
   {\int\limits_{-{\sqrt{\theta^2_c-\theta^2_x}}}
    ^{\sqrt{\theta^2_{c}-\theta^2_x}}}  \ \
     \frac{\theta^2_x \cos^{2} 2\rm\theta_B +\theta^2_y}
    {(\theta^2_x+\theta^2_y +\theta^2_{ph})^2} d\theta_y=
   $$
   $$
   =\frac{\theta^2_{ph}+\theta^2_x(1+\cos^{2}2\rm\theta_B)}
   {(\theta^2_{ph} + \theta^2_x)^{3\over2}} 
   \arctan{\sqrt{\frac{\theta^2_c-\theta^2_x}{\theta^2_{ph}+
   \theta^2_x}}}-
   $$
   $$
   -{{\sqrt{\theta^2_c-\theta^2_x}(\theta^2_{ph}
   +\theta^2_{x}\sin^{2}2\rm\theta_B)}
   \over(\theta^2_{ph}+\theta^2_c)(\theta^2_{ph}+
   \theta^2_x)}
   $$
   For the case where $\theta_c << \theta_{ph}$ we can write 
   a simpler expression:
        $$ 
       \Lambda_c(\theta_x)\approx{{\sqrt{\theta^2_c- 
       \theta^2_x}\over\displaystyle\theta^2_{ph}}{{\theta^2_c-
       \theta^2_{x}\cos2\theta_d}\over\displaystyle\theta^2_{ph}}}\eqno(18)
       $$
       After substituting $\theta_x = - u$ into the above we get the
       spectral line.

     Let us analyse the influence of mosaic structure on the
    spectral lineshape. In order to get the spectral distribution
    of the PXR beam in the aperture $\Delta\theta_x \sim 
    \theta_{ph}, \ \Delta\theta_y  >> \theta_{ph}$ 
    we have to use the following     distribution:
      $$
     \frac{dN}{du} = 
     \int\limits_{\Delta\Omega} d\theta_{x}d\theta_y\int\limits
     d\alpha_yF_m(\theta_x+  u, \ \alpha_y)\times
     $$
     $$
      \times \Lambda(\theta_x+2u, \theta_y + 2\alpha_y\sin\rm\theta_B)\approx
      {\int\limits_{\Delta\theta_x}d\theta_x F_m(\theta_x+u)}\cdot
     $$
     $$
     {\pi\over 2} {{\theta^2_{ph}+(\theta_x+2u)^2(1+\cos^{2}2\rm\theta_B)}
     \over{[\theta^2_{ph} +(\theta_x+2u)^2]^{3\over2}}},\eqno(19)
     $$

     To calculate the PXR spectrum emitted in the given angle 
      $\displaystyle \theta_x $ ( near $\omega_0 = 
      \rm\omega_B(1-\cot\rm\theta_B\theta_x)$),
     following Eq.(19) let us make a replacement, then the integrand
     in Eq.(19) will correspond to the spectrum sought for:
     $$
     {\partial^2\widetilde {N}\over\partial\theta_x\partial{u}}={{1\over
     \sqrt{2\pi}\sigma_m}}{\exp\{-{(\theta_x+u)^2\over2\sigma^2_m}\}}
     $$
     $$
     {\pi\over2}{{\theta^2_{ph}+(\theta_x+2u)^2{(1+\cos^{2}2\rm\theta_B)}}
     \over\lbrack\theta^2_{ph}+(\theta_x+2u)^2\rbrack^{3/2}},\eqno(20) 
     $$
     Upon integrating (20) within aperture one may get the shape of
     PXR spectral line, (e.g.see Fig.3). Let 
     $\displaystyle\theta_B={\pi\over4}$  and 
     $\displaystyle\sigma_m << \theta_{ph}$. Then
     the energy range  $\displaystyle\vert\theta_x+ u\vert << 
     2\sigma<<\theta_{ph}$
     and,therefore,
     $$
     {\partial^2\widetilde N\over\partial\theta_x\partial{u}}
     \approx {1\over2\sigma_m} {\sqrt{\pi\over2}} 
     {1\over\theta_{ph}}
     {\exp\{-{(\theta_x+u)^2\over2\sigma^2}\}}
     $$
     The distribution obtained has a maximum at $\displaystyle u_0=
     -\theta_x$,  i.e. at
     $$
      \omega_0=\rm\omega_B(1+\theta_{x}\cot\rm\theta_B).
     $$
     Contrary to the above, at $\displaystyle\sigma_{m} >> \theta_{ph}$,	
     position of the maximum is determined by the last term in Eq.(20):
     $$
     u_0= -{\theta_x\over2}, \ \  \omega_0=\rm\omega_B(1-{\theta_x\over2}
     \cot\rm\theta_B)
     $$
     and the spectral distribution has the form:
     $$
     {\partial^2\widetilde N\over\partial\theta_x\partial{u}}
      \approx{1\over2\sigma_m} {\sqrt{\pi\over2}}
     {\exp\lbrace-{\displaystyle\theta^2_x\over4\sigma^2_m}\rbrace}	
     {1\over\sqrt{\theta^2_{ph}+(\theta_x+2u)^2}}
     $$
     In this case the energy of PXR photons is defined by the
     elements of mosaic structure with 
     $\displaystyle\alpha_x=\theta_x+u \approx
     {\displaystyle 1\over2}\theta_x$
     (but not by the mosaic structure with $\alpha$ = 0),
     which results in an effective change of the angle 
     $\displaystyle\widetilde {\theta_d}$:
     $$
     \widetilde \theta_d=\theta_d+\theta_x
     $$
     Summing up the above considerations we may state that the main
     contribution into radiation along $\theta_x$  comes from the 
     mosaic  structure elements whose direction coincides  with 
     the Bragg direction.

     The approach described in previous chapter is a good 
     approximation for the case where the PXR line is located far 
     from the absorption edges, where the absorption length changes 
     in a drop. In order to take the $K$-edge influence into
     account it is necessary to calculate the spectral distribution
     $\displaystyle dN\over\displaystyle d\omega$  according to the method 
     described in the  previous chapter and then count the 
     distributions of photons with the energy above and below 
     the $K$-edge. These two groups of photons are absorbed in 
     the crystal in a different manner, i.e. they have different 
     length $ L_a $. Using,
     for each of the groups, a formula similar to (9) and then
     finding the sum, we may obtain the photon 	yield near the 
     $K$-edge of absorption.

     6. In conclusion we would like to note the following:

         a) beam divergence and crystal mosaicity  effect the PXR
      characteristics in a different manner;

         b) with increasing beam divergence (or mosaicity) the 
      well-known two-lobe angular distribution of PXR transform 
      into a  single-lobe distribution with maximum near Bragg direction;

         c) the spectral line width of PXR is defined by the collimator's
      aperture  (or detector's) and mosaicity, but not the divergence
      (or multiple scattering);

         d) and finally that the approach developed  makes it 
     possible to calculate ( to an accuracy where the  PXR kinematic 
     theory is true) all the PXR characteristics measured without
     addition of any other phenomenological parameters.

     The author thanks L.V.Puzyrevitch, T.D.Litvinova and C.Yu.Amosov 
     for the help with design of the paper.

\newpage

 {\Large \bf    Figure Captions} \vspace{4mm}

 Fig. 1. PXR angular distributions for 
            $\displaystyle\rm\theta_{B} = \pi/4$: \\
            1 - ideal case; \   2 - convolution with beam divergence 
            $\displaystyle\sigma_e=\theta_{ph}$; \\
            3 - Feranchuk-Ivashin model  for 
            $\displaystyle\theta^2_{ph}=\gamma^{-2} + 
            \frac{\omega^2_p}{\omega^2}+ \sigma^2 , \
            \sigma=\theta_{ph}.$ 

 \vspace{3mm}
 Fig. 2. PXR angular distributions for different mosaicities : \\
          1 - $\displaystyle\sigma_m = 0.$; \ 
          2 - $\displaystyle\sigma_m =   \theta_{ph}$; \ 
          3 - $\displaystyle\sigma_m = 2\theta_{ph}$. 
                
\vspace{3mm}
 Fig. 3. PXR spectral distributions for $\rm\theta_{B} =
         \pi/4 $ and $\Delta\theta_x = \pm 0.5\theta_{ph}$ \\
         Curve 1 - $\sigma_m = 0.2\theta_{ph},$ \
         Curve 2 - $\sigma_m = 4\theta_{ph}$.

\newpage
\vspace{2cm}
\unitlength=1cm
\begin{picture}(18,30)
\put(2,21.5){\epsfxsize=10cm\epsfbox{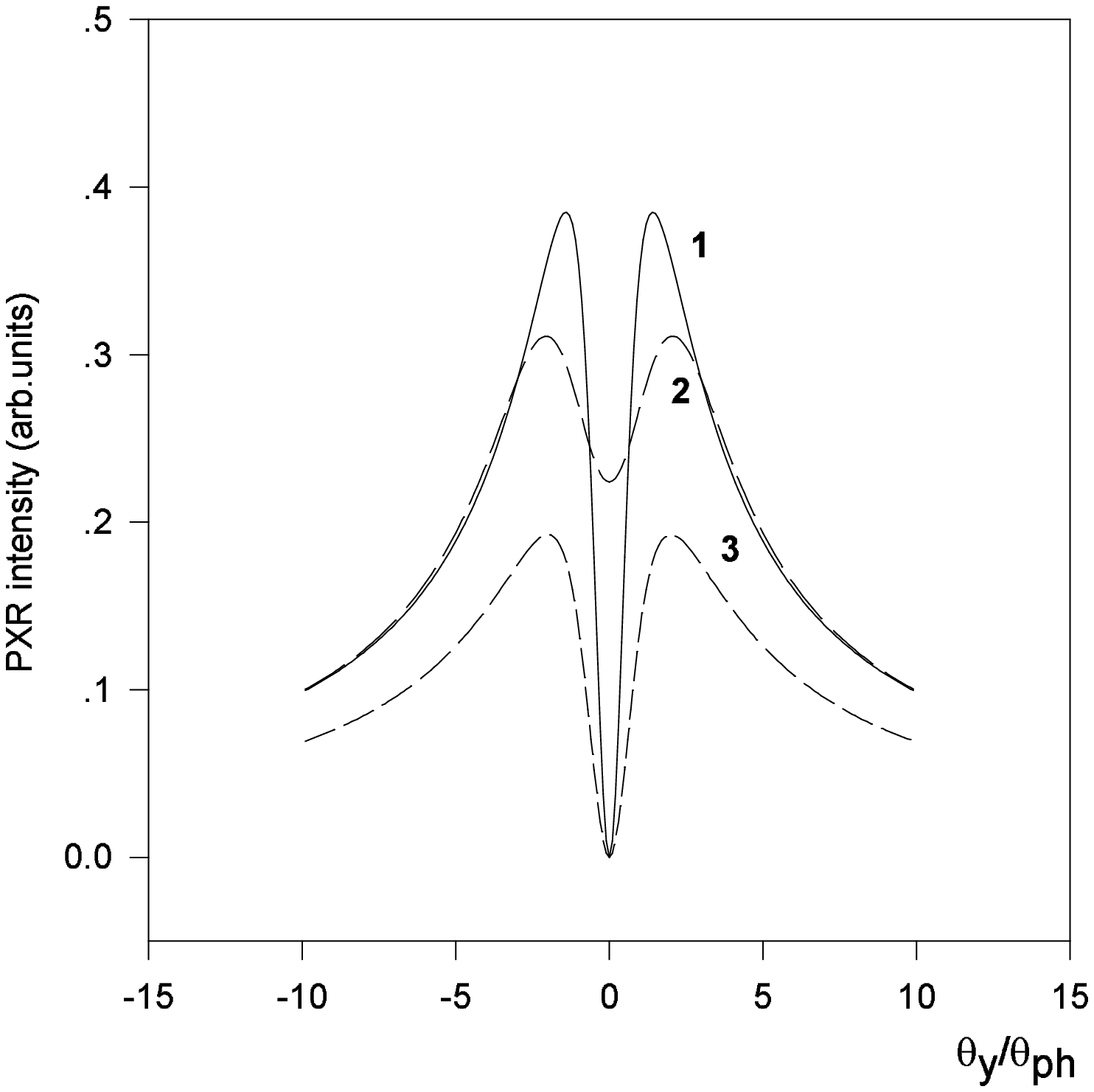}}
\put(6,24.2){\makebox{Fig. 1}}
\put(2,9.5){\epsfxsize=10cm\epsfbox{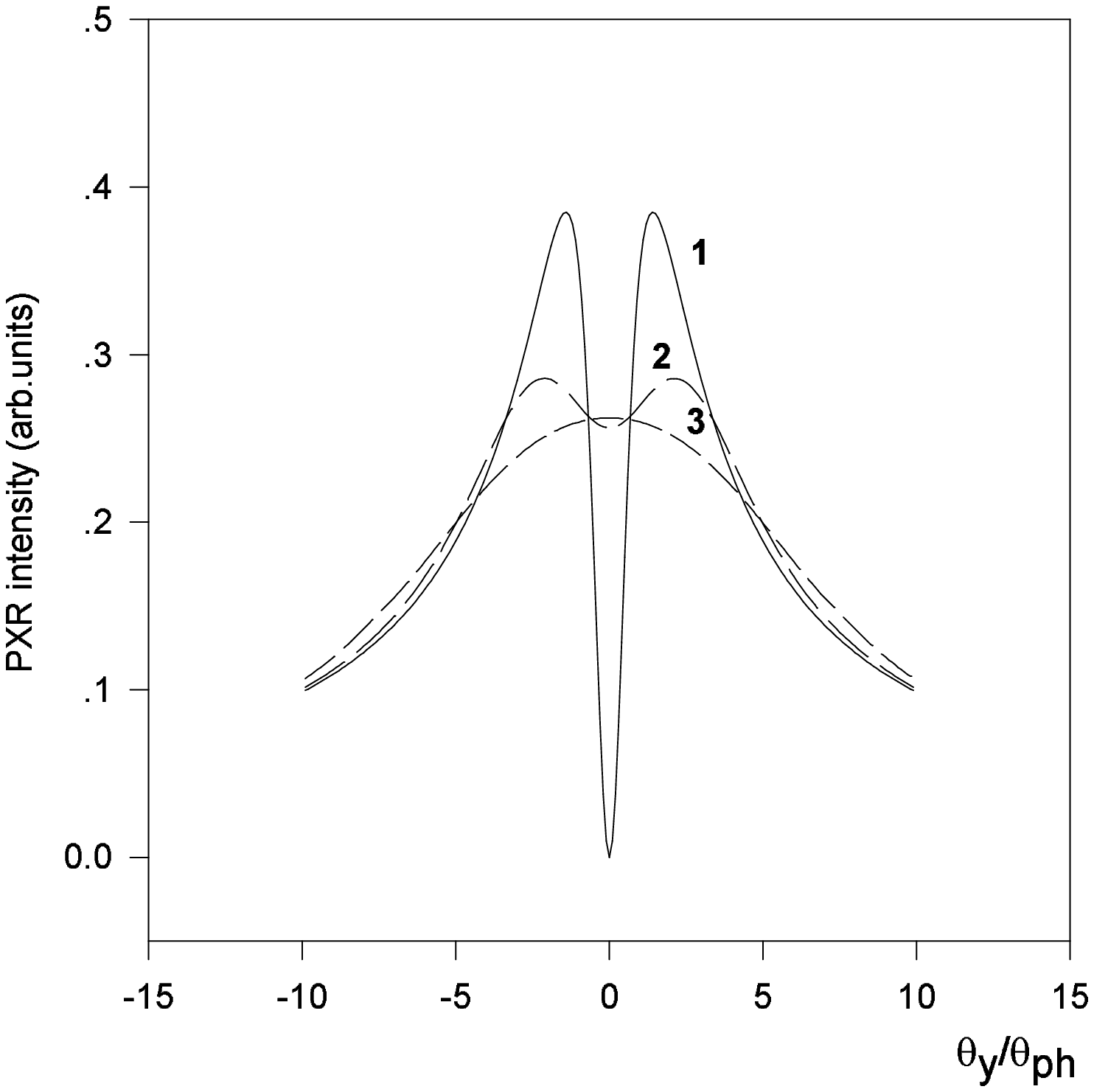}}
\put(6,12.1){\makebox{Fig. 2}}
\end{picture}

\newpage
\unitlength=1cm
\begin{picture}(18,30)
\put(2,13){\epsfxsize=10cm\epsfbox{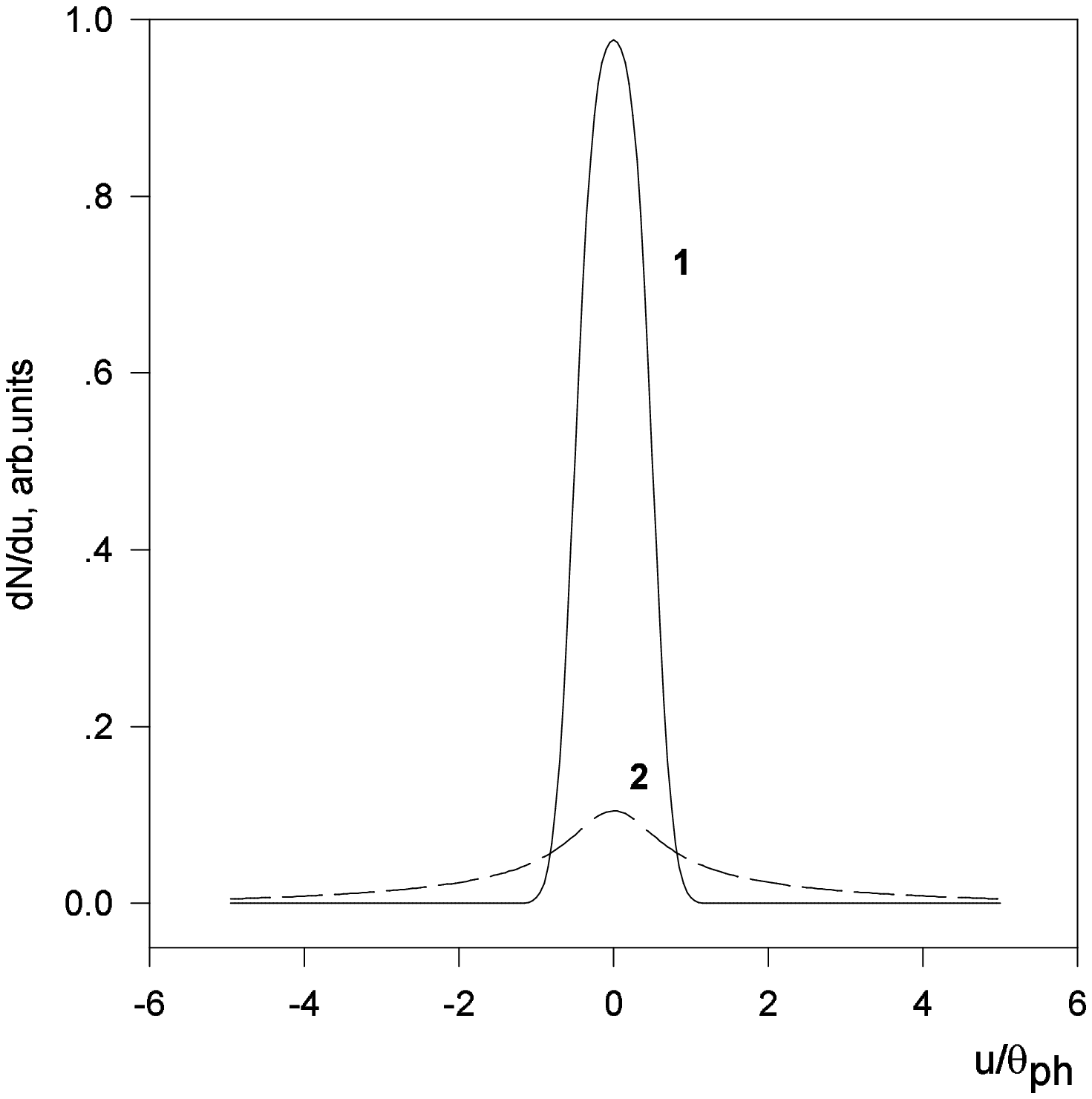}}
\put(6,15.5){\makebox{Fig. 3}}
\end{picture}

\end{document}